\documentclass[prb,reprint,groupedaddress,showpacs, showkeys]{revtex4-1}

\usepackage{graphicx}% Include figure files
\usepackage{dcolumn}% Align table columns on decimal point
\usepackage{bm}% bold math
\usepackage{booktabs}
\usepackage[table,xcdraw]{xcolor}
\usepackage{colortbl}
\usepackage[utf8]{inputenc}
\usepackage[T1]{fontenc}
\usepackage{mathptmx}
\usepackage{etoolbox}

\usepackage{graphicx}    % include figure files
\usepackage{dcolumn}     % align table columns on decimal point
\usepackage{bm}          % bold math
\usepackage{amsmath}     % extended math
\usepackage{amssymb}     % extended symbol set
\usepackage{multirow}    % table formatting
\usepackage{color}       % colored annotations
\usepackage{soul}
\usepackage{enumerate}

\makeatletter
\def\@email#1#2{%
 \endgroup
 \patchcmd{\titleblock@produce}
  {\frontmatter@RRAPformat}
  {\frontmatter@RRAPformat{\produce@RRAP{*#1\href{mailto:#2}{#2}}}\frontmatter@RRAPformat}
  {}{}
}%

\newcommand{\Fig} [1]    {Fig.~\ref{#1}}
\newcommand{\ie}         {\textit{i.e.,} }       % idem
\newcommand{\rhs}      {r.h.s.}               % include as {\rhs}

\newcommand{\dt}					{\Delta t}
\newcommand{\kBT}					{k_B T}
\newcommand{\kB}					{k_B }

\newcommand{\nm}					{\mathrm{nm}}

\newcommand{\epsLJ}				{\epsilon_\textrm{LJ}}

\newcommand{\Apull}				{A$^\textrm{p}$}

\newcommand{\distance}			{\mathbf{r}}

\newcommand{\FreeEne}				{\mathcal{A}}
\newcommand{\force}				 {F}

\newcommand{\ForceFree}[1]			{\bm{\mathcal{\force}}^{#1}}

\newcommand{\genCoord}			{Q}
\newcommand{\GenCoord}			{\mathbf{\genCoord}}

\newcommand{\mobility}				{\mu}

\newcommand{\MobilityGen}			{\boldsymbol{\mobility}^\genCoord}

\newcommand{\posCoM}				{R}
\newcommand{\PosCoM}				{\mathbf{\posCoM}}
\newcommand{\potential}				{\Phi}

\newcommand{\quat}				{q}
\newcommand{\Quat}				{\mathbf{\quat}}

\newcommand{\rand}				{\Theta}

\newcommand{\Rand}[1]				{\boldsymbol{\rand}^{#1}}

\newcommand{\Rg}					{R_g}

\newcommand{\tpull}        {\tau^\text{d}}
\newcommand{\trelax}        {\tau^\text{r}}

\newcommand{\MPtwenty}  {\textrm{MP}$_\textrm{20}$}
\newcommand{\MPthirtysix}  {\textrm{MP}$_\textrm{36}$}

\newcommand{\MPfortyeight}  {\textrm{MP}$_\textrm{48}$}
\newcommand{\MPsixty}  {\textrm{MP}$_\textrm{60}$}
\newcommand{\MPseventytwo}  {\textrm{MP}$_\textrm{72}$}
\newcommand{\MPnintysix}  {\textrm{MP}$_\textrm{96}$}
\newcommand{\MPoneeighty}  {\textrm{MP}$_\textrm{180}$}
\newcommand{\MPtwoforty}  {\textrm{MP}$_\textrm{240}$}

\makeatother
\begin{document}

\preprint{Paesani}

\title[Computational engineering of metaparticles]{
MetaParticles:
Computationally engineered nanomaterials with tunable and responsive properties}
% Force line breaks with \\
\author{Massimiliano Paesani$^{1,2,3}$}
\author{Ioana M. Ilie$^{1,2,3}$}%
 \email[Corresponding author: ]{i.m.ilie@uva.nl}
\affiliation{ 
$^1$ Van 't Hoff Institute for Molecular Sciences, University of Amsterdam, Amsterdam,
 The Netherlands \\ 
$^2$ Amsterdam Center for Multiscale Modeling (ACMM), University of Amsterdam, the Netherlands \\
$^3$ Computational Soft Matter (CSM), University of Amsterdam, the Netherlands
}%Lines break automatically or can be forced with \\

\begin{abstract}
In simulations, particles are traditionally treated as rigid
    platforms with variable sizes, shapes and interaction parameters.
While this representation is applicable for rigid core platforms, 
    particles consisting of soft platforms 
    (e.g. micelles, polymers, elastomers, lipids)
    inevitably deform
    upon application of external stress.   
We introduce a generic model for flexible particles which we call 
    \textit{MetaParticles} (MP).
These particles have tunable properties, can respond to applied tension
    and can deform.
A metaparticle is represented as 
    a collection of Lennard-Jones beads interconnected 
    by spring-like potentials. 
We model a series of metaparticles of variable sizes and symmetries, which
    we subject to external stress followed by relaxation upon stress release.
The positions and the orientations of the individual beads 
    are propagated by Brownian dynamics.
The simulations show that the mechanical properties of the metaparticles
    vary with size, bead arrangement and area of applied stress,
    and share an elastomer-like response to applied stress.
Furthermore, metaparticles deform following different mechanisms, \ie small
    MPs change shape in one step, while larger ones follow a
    multi-step deformation pathway, with internal rearrangements of the beads.
This model is the first step towards the development and understanding of 
    particles with adaptable properties with biomedical applications 
    and in the design of bioinspired metamaterials.
\end{abstract}

\maketitle

\section{\label{Sec:Intro}INTRODUCTION}
Nanoparticles (NPs) are versatile platforms in a variety of fields, 
    including biomedical application \cite{Mitchell:Nature_20_1474_2020, Wang:PR_114_56_2016, Khurana:BP_111_802_2019}
    cosmetics \cite{Drno2019,Lu2015}, 
    material engineering \cite{Bao2015} and the food industry\cite{Kumar2020}.
In cosmetics, nanoparticles %, in virtue of their sizes, 
    are used as filters of UV radiation in sunscreens or to enhance the delivery of active ingredients in dermal tissues\cite{Drno2019, Lu2015}.  
%Moreover, their antibacterial capabilities make them 
%    an effective addiction to bathing products and toothpastes\cite{Fattahi-Dolatabadi:GMS_2013, Prabhu:INL_2_2012}. 
In the food industry, nanoparticles are used as
    transport platforms of 'nutraceuticals' 
    towards specific tissues
    and for preservation purposes\cite{Huang:JFC_75_1750_2010, Bajpai:JFDA_26_1021_2018}.  
In the biomedical field, nanosized objects loaded with 
    molecules are commonly used as carriers to deliver 
    their content (e.g. nutrients, drugs, imaging agents) to a desired target \cite{Barua:NANO_9_1748_2014, Beach:chemrev_124_1520_2024}.  
Their diversity in size, functionalization and available platforms, 
    make nanoparticles excellent candidates
    for therapeutic activity in precision medicine \cite{Khurana:BP_111_802_2019, Sharma:NanoBiotech_20_1477_2022}. 
Medical applications of NPs range from cancer therapy, 
    where nanocarriers are used to actively target 
    cancerous cells\cite{Zheleznyak:WIRE_19_1939_2018,Sun:STT_8_2059_2023}
    to immunotherapy\cite{Guerrini:NatureBiotech_17_1748_2022}
    and genome editing where inhalable NPs formulations
    are used as therapeutic agents in cystic fibrosis\cite{Velino:Frontiers_7_2296_2019}.
Metal–organic nanocages are gaining momentum in the biomedical context,
    showing reduced cytotoxicity, good \textit{in vivo} biodistribution, and anti-cancer activities
    \cite{Bobylev:chemsci_12_7696_2021, Bobylev:chemsci_14_6943_2023}
%Given the complexity of the biological system, a nanocarrier should     fullfill a series of properties such as good diffusivity,    high bioavailability, specificity and excellent membrane passage abilities    \cite{intro2, more citations,natureintro}.
Nanoparticles are good candidates in the 
    rational design of novel materials
    with mechanical properties that are inaccessible 
    to ordinary materials, \ie metamaterials \cite{Kadic:natrevphys_1_198_2019,Tzarouchis:epjam_11_2272_2024}.
Nanoparticles have been used to 
    reprogram the deformation behavior of mechanical metamaterials
    \cite{Zheng:amt_29_101662_2022} and to 
    engineer stretchable materials with a broad range of organic or inorganic functionality \cite{Pham:advmat_25_6703_20132013} as well as biomedical 
    applications \cite{Tzarouchis:epjam_11_2272_2024}.
    
Nanoparticles consist of a core platform functionalized 
    with ligands on the surface.
These ligands are designed to enhance and modulate the
    individual and/or collective properties of NPs
    in specific environments \cite{Liu:NC_12_2041_2021}.
For instance, DNA-patched nanoparticles have been synthesized
    to control the self-assembly of colloidal metamaterials 
    with functionality-explicit architectures
    \cite{Liang:jacsAu_3_1176_2023}.
In the biomedical field, various nanoparticle cores 
    are designed to fit specific applications \cite{Mitchell:Nature_20_1474_2020}.
Hence, lipid nanoparticles are frequently used for 
    drug delivery purposes\cite{FonsecaSantos:IJN_1178_2015, Sercombe:Frontiers_6_1663_2015},
    because of their high stability, biocompatibility and specificity in
    the cellular environment.
While also improving drug bioavailability
    \cite{Patra:JNb_16_1477_2018, Volpatti:ACS_14_1936_2019}, 
    polymeric nanoparticles struggle with poor stability and 
    solubility due to their hydrophobic nature, which often
    limits their drug delivery abilities \cite{Rideau:RSC_47_1460_2018}.
Inorganic nanoparticles consisting of metallic cores are abundantly 
    used for imaging purposes \cite{Han:RSC_11_2040_2019}.
Their optical and electrical properties make them suitable 
    for specific delivery conditions, 
    like thermal-based therapeutics \cite{Wang:ACS_5_2470_2019} or, 
    more recently, \textit{in vivo} diagnostics \cite{Yang:IJP_625_122122_2022, Wagner:Elsevier_94_1742_2019}.
    
Computer simulations complement experimental research
    to provide insight into specific chemical and physical
    processes, e.g. informed ligand design, property prediction, mechanisms in biological environments.
The aid in the development of 
    biological-compatible nano-structures to guide and inform on processes
    exceeding experimental resolution
 %   where experimental resources are limited or time consuming
    \cite{T:ML_368_0167_2022,Casalini:Frontiers_2019,Mahmood:ACS2022}. 
At atomistic resolution, molecular dynamics simulations
    provide insight into the fine details 
    of coated nanoparticles \cite{Habibollahi:Elsiver2023}.
They are powerful tools to suggest and investigate
    the effects of specific chemical modification on the biological
    environment, e.g. interactions with membranes \cite{Zhang:NPG2021},
    behavior in the blood flow \cite{Yamaguchi:Springer2010} or even make
    predictions that can be subsequently
    experimentally tested \cite{Promkatkaew:ACS2024}.
Gold nanoparticles are amongst the most studied candidates
    due to their size (< 10~nm), optimized force-field parameters and
    ease in introducing ligand modifications \cite{Milano:ACS_115_1932_2011, Perfilieva2018, Lee2009}.
However, atomistic detail comes with a computational cost, particularly
    when studying accumulations of nanoparticles or biological response, 
    which require access to long time- and lengthscales (e.g. membrane
    associated processes, diffusion).

\begin{figure*}
\includegraphics[width=\textwidth]{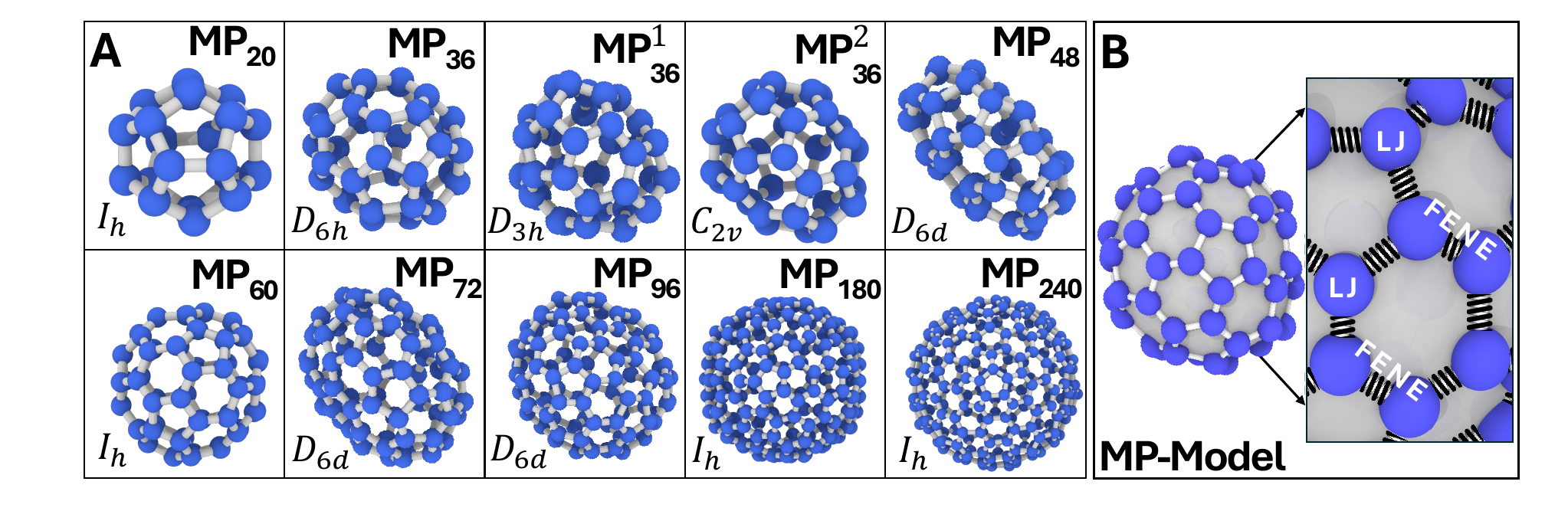}% Here is how to import EPS art
\caption{\textbf{Coarse-grained models of the metaparticles}. 
    Shown are snapshots of the topologies 
    and point symmetries
    of the ten developed metaparticles consisting of 
    20 to 240 beads (A) and a schematic representation of
    the novel metaparticles model described by
    Lennard-Jones (LJ) interactions and FENE bonds 
    between the constitutent beads (B). 
    $\textrm{I}_h$ - icosahedral symmetry; 
    D$_\textrm{6h}$ - symmetry which features a 6-fold rotational axis with the inclusion of horizontal reflection planes;
    D$_\textrm{6d}$ - symmetry which features a 6-fold rotational axis with the inclusion of diagonal reflection planes;
    D$_\textrm{3h}$ - trihedral symmetry with 3-fold rotational axis;
    C$_\textrm{2v}$ - 2-fold rotational symmetry with the inclusion of vertical reflection planes.
\label{Fig:fullerene}}
\end{figure*}
Coarse-grained models overcome these difficulties by 
    grouping atoms into a single beads, thereby
    reducing the degrees of freedom of a system and making it computationally more affordable when simulating large systems \cite{Izvekov:ACS_109_1520_2005,Zhang:ACS_9_1936_2015}.
In particular, nanoparticles designed as single beads 
    decorated with interaction patches to mimic the surface moieties 
    are ideal to explore the influence
    of size and shape \cite{Dasgupta:nanolet_14_687_2014,Vacha:nanolet_11_5391_2011, MartinezVeracoechea:ACS2011}, 
    ligand concentration and distribution 
    \cite{Schubertova:sm_11_2726_2015,Gkeka:jpcl_4_1907_2013}
    to investigate cellular uptake and
    accumulation \cite{ZurawWeston:nanoscale_11_18464_2019}.
These minimal coarse-grained models are suitable candidates to help with 
    gaining insight into 
    generic mechanisms while reaching timescales comparable to experiments. 
Most commonly, they use rigid representations of the
    nanoparticle core, while in reality 
    their intrinsic flexibility arising from the core properties
    might affect 
    their morphology and their ability to accurately capture environmental effects 
    or overcome biological barriers \cite{Shae:NN_14_1748_2019}.
Hence, though relevant, 
    the mechanical properties of nanoparticles,
    which influence their properties at individual
    and collective level, are often neglected in
    minimal models. 
For instance, both experimental \cite{Li:DD2017} 
    and computational studies \cite{Li:Nanoscale_7_2040_2015,Midya:ACS_17_1936_2023}
    have shown that soft NPs can enable drug delivery 
    through mechanisms that avoid endocytosis, 
    a process requiring biological membrane curvature and breakage, 
    which slows down absorption and introduces significant complications \cite{Lacy:Wiley2018,Stachowiak:NCB2013}. 
This process reminds of internalization viral mechanisms 
    that leverage intrinsic flexibility for more rapid cellular entry 
%    which computational models have also attempted to recreate
    \cite{Kobile:Nucleus2012,Harrison:Virology_479_0042_2015}.
In this context, tailoring physical properties of individual
    nanoparticles to adapt to the environment
    can aid in the development of 
    multilayer metamaterials \cite{Tang:acsnano_14_14895_2020}.
Despite the substantial evidence linking
    mechanical properties to
    functionality and macroscopic properties of nanoparticles, 
    existing coarse-grained model do not take into 
    consideration the intrinsic flexibility of 
    a nanoparticle. 
    
Here, we develop a coarse grained model able to capture 
    the intrinsic flexibility of nanoparticles, arising
    from soft core platforms or anisotropic ligand coverage.
We introduce a model, in which a nanoparticle is
    represented as a set of hard 
    spheres interconnected with flexible bonds 
    that can respond to applied external tension, hence adapting
    to the environment.
Due to their modularity, versatility and adaptability, we 
    refer to the designed
    nanoparticles as metaparticles (MPs).
The topologies of the metaparticles rely on 
    symmetrical structures
    inspired by the structures of 
    fullerenes (\Fig{Fig:fullerene}). 
Our results show that the metaparticles have 
    response to stress that is similar to the one of elastomers.
Additionally, the response it modulated by the delicate balance between
    size, topology and area of applied stress.
Lastly, the deformation and recovery timescales of large metaparticles 
    depend on the mechanisms of deformability and relaxation.

%%%%%%%%%%%%%%%%%%%%%%%%%%%%%%%%%%%%%%%%%%%%

\section{Model and Methods}
\subsection{Metaparticle model}
The model developed here is a generic representation
    of a particle with responsive and adaptable 
    properties, which we refer to as a \textit{MetaParticle} (MP).
We seek to mimic the mechanical properties
    of nanoparticles with soft cores such as 
     nanogels \cite{Soni:Elsevier2016}, 
     polymers \cite{landel1993mechanical}, 
     nanoparticles made from dendrimers as building blocks \cite{vanDongen:RSC2013}, micelles \cite{Ghezzi:Elsevier2021}, vesicles \cite{Herrmann:Nature2021} or biomolecular condensates \cite{Banani:Nature2017} and elastomeric materials with characteristics similar to biological tissues\cite{Chen:Elsevier2013}.
To achieve this, we introduce a generalized model of a metaparticle
    consisting of multiple interconnected beads, 
    symmetrically arranged in topologies inspired by 
    the structures of fullerenes.
These are symmetric  
    nanostructures that consist of carbon atoms interconnected
    by single or double bonds to form closed or semi-closed cage-like molecules.
For our model, we consider only the closed topologies.
These molecules form an interconnected network of 
    12 pentagons 
    and a variable number of hexagons or, less common, heptagons \cite{Schwerdtfeger:WIRE_5_1759_2014}, resulting in 
    hollow structures (e.g. spheres, ellipsoids, spherocylinders) 
    (\Fig{Fig:fullerene}A). 
%Chemically, each carbon atom is connected to three other
%    carbon atoms (and a hydrogen atom).
Topologically, this arrangement translates into a network of 
    vertices joined by three edges.
In our metaparticle model, each vertex is represented as a 
    hard spherical bead.
To introduce the flexibility element, each bond (or edge) 
    has a spring-like behavios (\Fig{Fig:fullerene}B).
To investigate the mechanical properties of MPs with
    comparable topologies, yet differing in size and
    arrangement, we chose ten model systems.
They range from small metaparticles with twenty beads
    in their composition 
    to larger MPs with 240 constituents.
 The smallest model, {\MPtwenty}, 
    consists only of twelve pentagons, while all the other models have, 
    at least, one hexagon in their topology. 
To account for variability in the shape and topology,
    we consider metaparticles with different symmetries (\Fig{Fig:fullerene}).
Furthermore, to characterize systems with the same number of beads
    but different topologies, we use 
    three model systems with different 
    symmetries for the MPs with
    36 vertices. 
\subsection{Interaction potentials}
In our model, the metaparticle consists of a network of 
    interconnected vertices represented as beads.
The non-bonded interactions between two beads
    in the metaparticle composition
    is given by a Lennard Jones potential,
\begin{equation}
\potential_\textrm{LJ} (\distance) =
\begin{cases}
    4 \epsLJ \left[\left(\frac{\sigma}{\distance}\right)^{12} - \left(\frac{\sigma}{\distance}\right)^6\right] & ~\textrm{if} ~\distance < 2.5\sigma \\
    0 & ~\textrm{if} ~\distance  \geq 2.5\sigma
\end{cases} 
\label{Eq:LJ}
\end{equation}
    where $\distance$ is the distance between the centres of mass of the beads,
    $\sigma$ the particle diameter and $\epsLJ$ the interaction strength.
To account for the connectivity between the particles, we represent
    the edges as elastic springs described by
    a finite extensible nonlinear elastic (FENE) potential usually employed for bead-spring polymer models\cite{Morthomas:APS2017, Wedgewood:NV_40_0377_1991,Zhou:JNF_116_0377_2004}.
The FENE potential is given by
\begin{equation}
\potential_\textrm{bond} (\distance)=-\frac{1}{2} k_\textrm{bond} r_0^2 \log \left[1-\left(\frac{\distance}{r_0}\right)^2\right] ~\textrm{for} ~ 0 < \distance < r_0
\label{eq:FENE}
\end{equation}
where $r_0$ is the maximum bond extension and $k_\textrm{bond}$ the spring constant of the potential.

%We set K at 30 $\epsilon$ /$\sigma^2$, similarly to what is done in the literature with polymers (XXX), and $R_0$ at 1.8 $\sigma$, differently to the common literature value of 1.5 $\sigma$. The goal of this change was to give the system more flexibility when under persistent and strong induced stress. The spring constant value of 30 is resistant enough to avoid chain crossing\cite{d51} and sufficiently small to allow us to employ larger time step 
%$\delta$$\tau$\cite{d50}.

%%%%%%%%%%%%%%%%%%%%%%%%%%%%%%%%%%%%%%%%%%%%%%%%

\subsection{Dynamics}
The simulations were performed using Brownian dynamics
    (BD), a method which implicitly accounts for solvent effects 
    in the equation of motion.
Briefly, the average effect of the collisions of the solute with the solvent
    are represented by a friction and the fluctuations around the friction
    are accounted for by a stochastic Markovian process.
Hence, in BD, the motion of a particle arises from the balance between a conservative force and the friction effect arising from the solvent. 
Here we present the generalised equation of motion
    as the expanded equations have been previously introduced
    \cite{Ilie:AIP_142_1089_2015,Ilie2,Delong:AIP2015} and implemented in LAMMPS
    \cite{Thompson:CPC_271_0010_2022}.
The generalized Brownian dynamics equation of motion \cite{Oettinger, Gardiner} reads as
	\begin{align}
	\label{Eq:GEoM}
		\GenCoord(t+\dt) - \GenCoord(t) 
		 = 
		&- \MobilityGen \frac{\partial \FreeEne}{\partial \GenCoord}\dt \nonumber \\
		&+ \kBT \frac{\partial}{\partial \GenCoord} \cdot \MobilityGen \dt \nonumber \\
		&+ \left( \MobilityGen \right )^{1/2} \Rand{\genCoord}(t) \sqrt{2\kBT \dt},
	\end{align}
where $\GenCoord$ represents the full set of generalised coordinates.
Depending on the equation of motion, \ie translational or rotational, 
    $\GenCoord$ may be replaced by the position of a particle, $\PosCoM$,
	or the quaternions $\Quat$ representing the orientation of a particle, respectively
    \cite{Ilie2,Ilie:AIP_142_1089_2015}.
The first term on the {\rhs} represents the displacement over a time step $\dt$ 
		which arises from the balance between
		a conservative force $\ForceFree{} = -\partial{\FreeEne}/\partial{\GenCoord}$ and the solvent friction,
		with $\FreeEne$ the free energy as a function of the generalised coordinates.
The second term on the {\rhs} represents the drift originating from the inhomogeneity of the 
    mobility tensor, which accounts for the recovery of the Boltzmann 
    probability distribution.
The last term on the {\rhs} represents the Brownian contribution and
	 $\Rand{\genCoord}$ is a three-dimensional time-dependent Markovian vector,
    where its components are chosen such that they have
	zero mean, unit variance, no correlations across them and no memory.

%This time integration updates, each timestep, velocity, positions and quaternion orientation of all particles in the system based on viscous and random forces in an attempt to reflect the physics of particles in fluid. Brownian dynamics algorithms were specifically developed with the aim of simulating rigid colloid’s motion in fluidic solutions and, differently from the similar Langevin Dynamics algorithms, they employ first order differential equation to propagate positions in time. Because of that, the choice of Brownian dynamics as a time integrator allows us to reach higher simulation time, due to larger timesteps, with relatively low computational cost. 
%Quaternions are a desirable choice when describing the orientation of a rigid body in three dimensions due to several reasons: they are computationally efficient, they allow for smooth interpolation between orientation and they can help with avoiding the Gimbal Lock, a problem associated with Euler angles where some orientations can cause a loss of rotational degrees of freedom\cite{q1}

%%%%%%%%%%%%%%%%%%%%%%%%%%%%%%%%%%%%%%%

\subsection{\label{sec:level2}Simulation details}
Throughout this work, we use reduced 
    Lennard–Jones units \cite{FrenkelSmit,AllenTildesley}
    with $\varepsilon$ and $\sigma$ representing the units of energy and length, respectively. 
Time unit is $\tau=\sigma/\sqrt{m/\varepsilon}$, where m
    is the particle mass and the temperature is given in $\varepsilon/\kB$ 
    ($\kB$ is the Boltzmann constant).
Each system was initialized by placing a metaparticle in a cubic box
    with period boundary conditions (40 $\sigma$ per edge).
%All simulations were carried out in implicit water with a viscosity of $10^{-3}$ Pa s.
The template conformations for the MPs were fullerene structures with
    twenty to 240 vertices\cite{Tomnek2014} (\Fig{Fig:fullerene}A). 
    %\textcolor{red}{extracted from nanotube.msu.edu/fullerene\cite{Tomnek2014}}
 
%The mobilities of single particles are readily calculated from the 
%    Stokes-Debye-Einstein equations for a sphere \cite{Koenderink:langmuir_16_5631_2000}.
The beads in the MP interact via 
    bonded and non-bonded interactions.
An individual particle interacts with its 
    neighbors via the elastic FENE potential, with a bond strength
     $k_{bond} = 30 \varepsilon/\sigma^2$ and a maximum extensibility $r_0=1.8\sigma$.
The maximum extensibility is slightly higher than in the 
    Kremer–Grest model \cite{Kremer:jcp_92_5057_1990}
    to account for a longer adaptability range of the metaparticle.
Additionally, individual beads can interact via a 
    generic hydrophobic potential
    described by a Lennard-Jones potential, with 
    an interaction strength $\epsLJ=1 \varepsilon$, and if 
    they are within a cut-off distance $r_c = 2.5\sigma$ from each other.    
The timestep was fixed at $\dt=10^{-4}\tau$, where $\tau$ is the BD time unit,
    and the systems were evolved using Brownian dynamics simulations 
    using the LAMMPS simulation package
    \cite{Ilie:AIP_142_1089_2015,Ilie2,Delong:AIP2015,Thompson:CPC_271_0010_2022}.
    
In a typical set of simulations, a metaparticle is first relaxed 
    for $10^5 \tau$ in absence of restraints and only
    being subjected to Brownian motion. 
Next, the MP is mechanically deformed by performing pulling simulations
    for $10^5 \tau$.
For this, a set of 12 forces 
    (10 for {\MPtwenty}) were applied on 
    the equilibrated structure.
Each force was applied on anti-parallel sets of beads, 
    \ie antipodal hexagons or pentagons.
To avoid shearing effects, each system was realigned with the 
    antipodal surfaces aligned along the direction of applied deformation. 
%We then calculated the end-to-end distance along the axial pull and the transverse end-to-end distance perpendicular to the axial one. Both distances were defined as the distance between the two centers of mass of the two chosen sets of particles at each timestep and their values were saved every 10 timesteps.
In the final step, the applied stress is removed and 
    the metaparticle is simulated
    in absence of restraints till each specific MPs is fully relaxed.

%%%%%%%%%%%%%%%%%%%%%%%%%%%%%%%%%%%%%%%%%%%%%%%%%%%%%%%
\section{Results}
\begin{figure}
\includegraphics[width=\columnwidth]{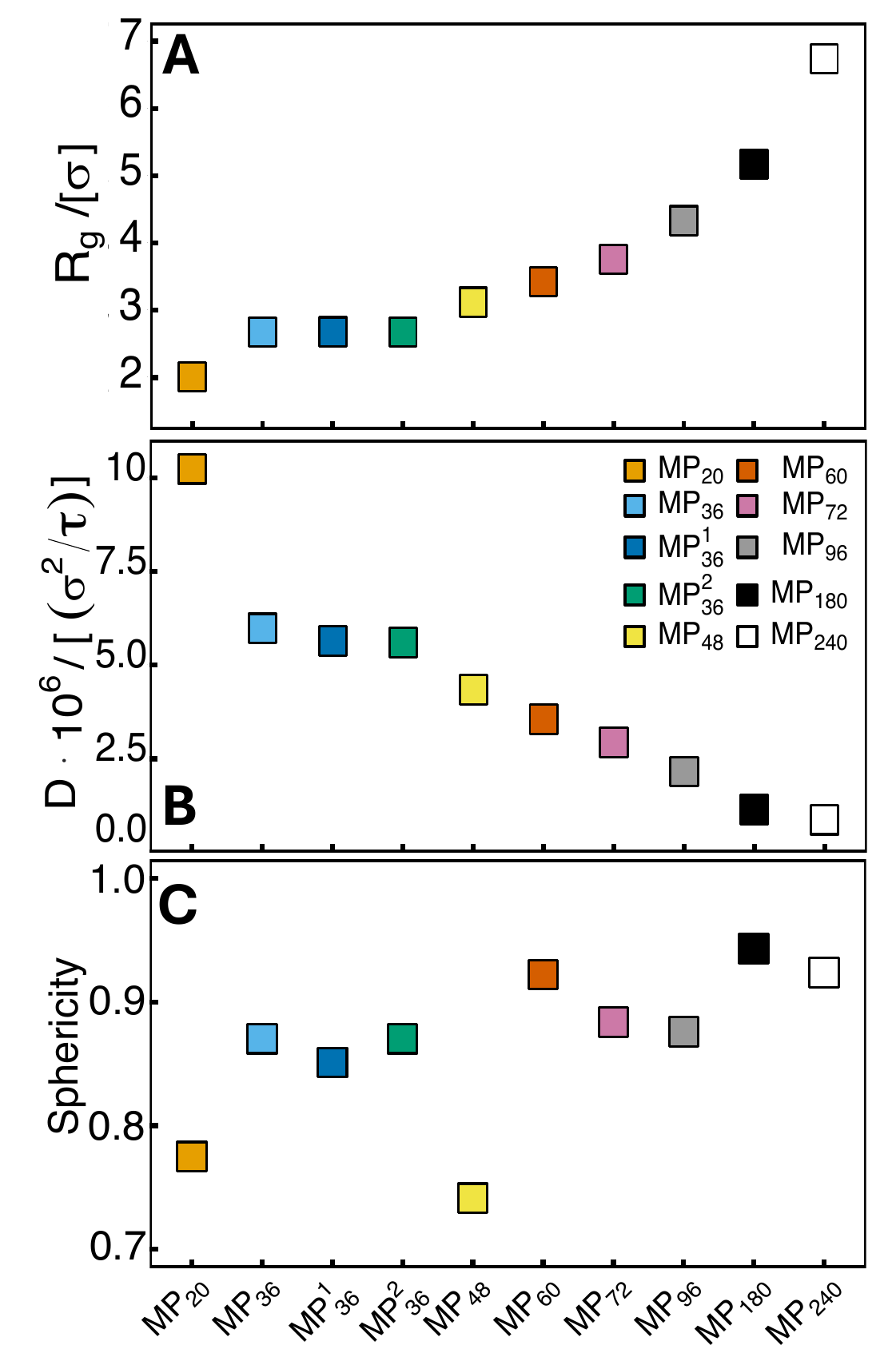}
\caption{\textbf{Properties of the metaparticles.}
    Radius of gyration, $\Rg$ (A), translational diffusion
    coefficients (B) and sphericity (C).
   %The error bars are small and 
    %represent the standard error calculated 
    %s the standard deviation of the mean. 
    The radii of gyration and the sphericities 
    are calculated on the equilibrated MPs. 
    %MPs with symmetry $D_6d$ show different behaviour from 
    %the expected size-dependent trend.
    \label{Fig:Rg}}
\end{figure}
\subsection{\label{sec:Diffusion}Metaparticle diffusion}
%To assess the diffusive behavior of the studied metaparticles, 
%    we simulated each MP for $10^5 \tau$
%    in absence of restraints using a time-step $\dt=10^{-4}\tau$.
First, we evaluated the radii of gyration $\Rg$ of the metaparticles
    as the ensemble average over the positions of
    all beads in an MP over time (\Fig{Fig:Rg}A).
We found that the $\Rg$ increases with the number of
    beads in the MPs, ranging from
    2.01 $\pm$ 0.006$\sigma$ for {\MPtwenty} to  6.74 $\pm$ 0.008$\sigma$ 
    for {\MPtwoforty}.
The {\MPthirtysix} variants are essentially identical in 
    their radii of gyration, despite the 
    different topologies, {\ie} 2.68 $\pm$ 0.007$\sigma$, suggesting that bead arrangement marginally affects
     metaparticle size.
    
Next, we investigated the diffusive behavior of the MPs
    by analyzing their mean squared displacement as
$\text{MSD} = \langle \left[ \PosCoM(t_0+\dt) - \PosCoM(t_0) \right]^2 \rangle = 6 \text{D} \dt + \mathcal{C}$.
The brackets $< >$ represent the average over all time 
    origins $t_0$ within the simulation run, 
    D the metaparticle translational diffusion coefficient 
    and $\mathcal{C}$ a constant.
In agreement with the $\Rg$ analysis, we find that 
    the translational diffusivity of the MPs
    reduces with increasing system size, \ie
    the smallest MP in diameter,
    {\MPtwenty} diffuses the fastest, while {\MPtwoforty} is
    the slowest (\Fig{Fig:Rg})B.
The three {\MPthirtysix} metaparticles have comparable 
    diffusion coefficients, showing that topology only
    marginally affects diffusion.

%%%%%%%%%%%%%%%%%%%%%%%%%%%%%%%%%%%%%%%%%%%%%%%%%%%%%%%%
\subsection{\label{Sec:Topo}Topological analysis}
By design, our metaparticles have symmetrical topologies 
    inspired by the structures
    of closed fullerenes (\Fig{Fig:fullerene}). 
Hence, each MP consists of 12 pentagons and a variable number of hexagons,
    and their arrangement is closely linked to shape and properties.
Pentagons induce curvature, which is essential to close the 
    metaparticle and form a spherical or quasi-spherical shape.
The MPs with less than sixty beads have pentagons 
    that share an edge with other pentagons, 
    with {\MPtwenty} presenting only pentagons in its structure.
Systems with more than sixty beads have pentagons surrounded by hexagons,
    following the isolated pentagon rule (IPR) rule\cite{Kroto1987}.

To gain insight into the topological effects on the behaviour of the MPs,
    we first analyzed their sphericity, \ie the resemblance to a perfect sphere.
The sphericity was determined from the ratio between the 
    volume of a metaparticle, ${\rm V}_{{\rm MP}}$, 
    and its surface area ${\rm A}_{{\rm MP}}$
    \cite{Aschenbrenner:jsr_26_1527_1956} 
    as ${36\pi\cdot{\rm V^2}_{{\rm MP}}\over {\rm A^3}_{{\rm MP}}}$ and
    following the isoperimetric quotient
    definition\cite{Schwerdtfeger:WIRE_5_1759_2014,Pisanski:ACS_37_0095_1997}.
${\rm V}_{{\rm MP}}$ and ${\rm A}_{{\rm MP}}$ were calculated using the alpha-shape 
    method\cite{Edelsbrunner:ACM_13_43_1994,Stukowski:JOM_66_299_2013}
    as implemented in Ovito\cite{ovito}.
Values of $\Psi$ close to 1 indicate nearly spherical objects, while
    decreasing values indicate deviations
    from a spherical shape with the object 
    becoming nearly flat as $\Psi$ approaches 0.

The results show that the larger metaparticle with icosahedral symmetry,
    \ie {\MPsixty}, {\MPoneeighty} and {\MPtwoforty}, 
    approach the most the spherical shape as this arrangement
    reduces the strain on the bonds and the angles, while seeking to
    minimize the surface area-to-volume ratio.
{\MPoneeighty} 
    has the highest resemblance to a sphere, given
    the homogeneous distribution of pentagons on its surface (\Fig{Fig:Rg}C).     
Compared to its smaller counterpart {\MPsixty} with the same symmetry, 
    {\MPoneeighty} consists of more beads, which allow for more favorable packing,
    even distribution of strain over the bonds and smoother curvature  
    thereby better approximating a sphere. 
The largest MP of the family, 
    {\MPtwoforty}, has a sphericity slightly
    lower than {\MPoneeighty}, 
    despite the higher number of beads and implicitly
    hexagons in its composition.
This is an inherent effect of the sphericity estimation using the isoperimetric
    quotient definition, which is consistent with previous calculations
    on fullerenes\cite{Schwerdtfeger:WIRE_5_1759_2014}.
Additionally, the metaparticles introduced here are flexible structures
    subjected to Brownian motion, and do not account 
    for electronic motion and chemical
    stability as in the case of real fullerenes.
%This further contributes to the deviations from the chemically synthesized stable fullerenes.
The MPs with D$_\textrm{6d}$ symmetry, \ie {\MPfortyeight}, {\MPseventytwo} 
    and {\MPnintysix} have lower sphericity than the 
    large icosahedral
    MPs.
Their sphericity correlates with size (\Fig{Fig:Rg}C).
Hence, the smaller {\MPfortyeight} has the lowest sphericity 
    due to its oblate arrangement, which has
    a lower number of hexagons in its composition, 
    leading to higher bond strains as compared to 
    {\MPseventytwo} and {\MPnintysix}.
The {\MPthirtysix} isoforms have comparable 
    sphericities (0.87, 0.85, 0.87) 
    independently of their symmetries.
Hence symmetry, bead packing and implicitly
    the distribution of hexagons do 
    not have a major impact on the
    sphericity of the small metaparticle. 
{\MPtwenty}, the smallest MP with 
    icosahedral symmetry and the outlier among the MPs
    has lower sphericity as compared to the
    {\MPthirtysix} isofroms.
This is due to its exclusive pentagonal composition, 
    which restricts efficient packing.
All in all, the topological analysis reveals that 
    metaparticles symmetries affect their sphericity
    to a larger extent than their size.

%%%%%%%%%%%%%%%%%%%%%%%%%%%%%%%%%%%%%%%%%%%%%%%%%%
\subsection{\label{sec:Mechanical}Mechanical properties}
To investigate the effects of size and topology on the mechanical properties 
    of the metaparticles, we studied the tensile stress response. 
Briefly, we deformed the individual structures by
    applying tensile forces on anti-parallel particle sets, \ie diametrically 
     opposed pentagons or hexagons.
First, each equilibrated MP was recentered in the simulation box
    and the centres of mass of the set of pulling area were 
    aligned along one of the principal axes to facilitate uniaxial deformation.
Then, a series of forces were sequentially applied on 
    diametrically opposed hexagons 
    (\ie previously aligned sets of particles), 
    except for {\MPtwenty}, which is devoid of hexagons.
We applied a total of 38 forces on each metaparticle
    (see SI for details),
    cumulating 38$\cdot 10^5$ $\tau$ sampling per system. 
The maximum force applied differs per system and 
    was chosen as the highest stress the MP
    can withstand prior to bond breakage.

\begin{figure}[h!]
\includegraphics[width=\columnwidth]{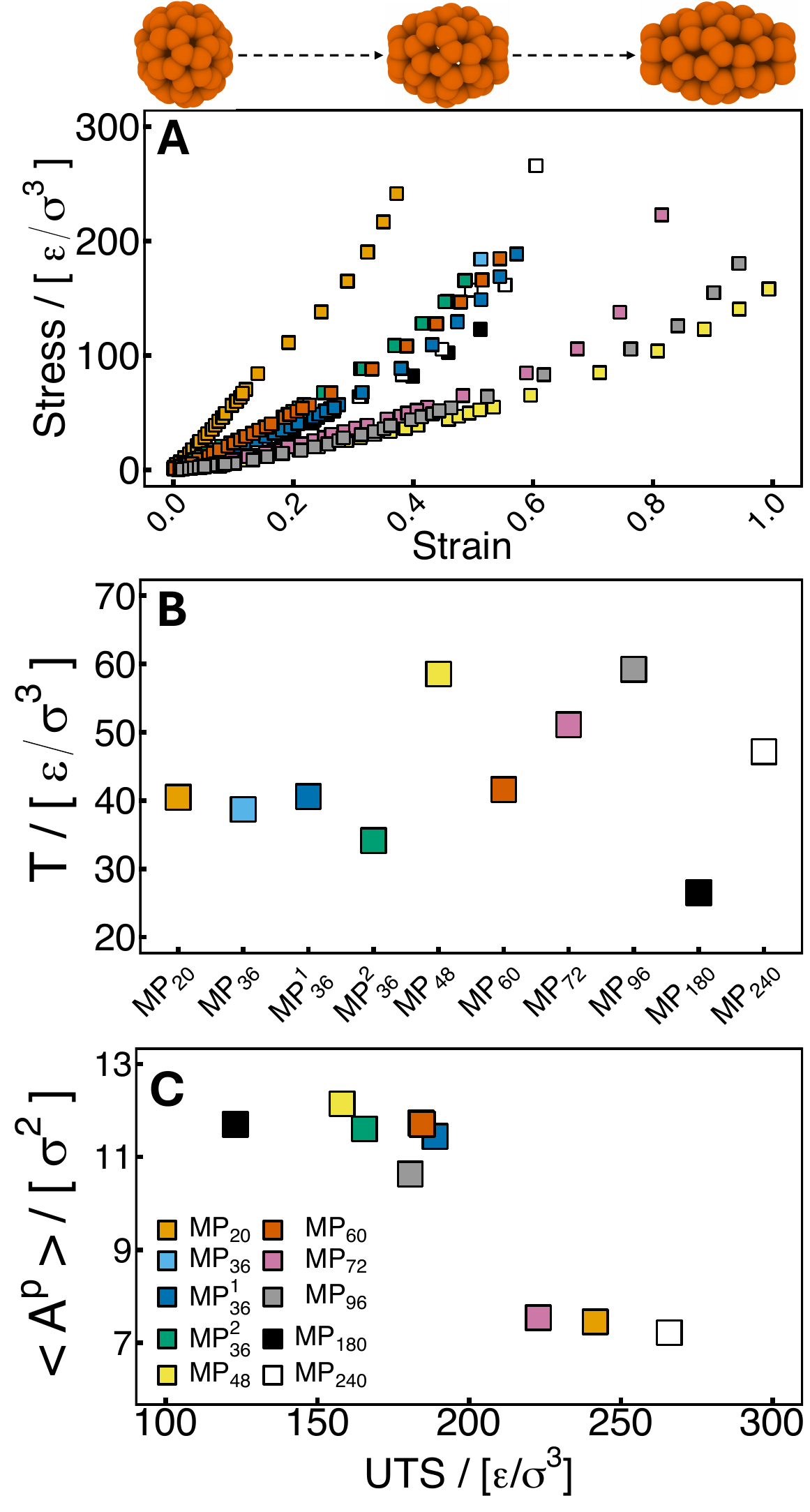}% Here is how to import EPS art
\caption{\label{Fig:stress} \textbf{Mechanical properties of the metaparticles.} Stress-strain curves. Highlighted are snapshots of {\MPsixty} at 5\%, 26\% and 55\% strain (A).
The strain is calculated as the change in axial length of the nanonparticle relative to 
the initial length.
Toughness calculated as the integration of the stress-strain curve (B).  Ultimate Tensile Strength (UTS) plotted against the average pulling area $A^p$ (C). The UTS is defined as maximum stress applied to the MPs.}
\end{figure}

%%%%%
The stress-strain curves have a non-linear response
    (\Fig{Fig:stress}A).
At low strains (< $\approx$ 5\%), 
    the stress response is linear and
    only marginal mechanical response of the 
    metaparticles is observed, with the thermal noise being dominant (\Fig{Fig:stress}A and Fig. S1).
%\textcolor{red}{idea to maybe have later a snapshot of the energies to show the distribution.}
At higher strains (> $\approx$ 5\%),
    the stress response shows an exponential behavior.
The bonds start to align in parallel to the direction
    of the applied stress and the MPs deform.
Notable, is that the metaparticles  heterogeneously resist the applied stress
    prior to bond breakage, \ie the maximum applied stress
    varies across the different MPs.
Furthermore, the stress-strain profiles vary with 
    metaparticles size and topology, and exhibit a 
    response similar to that of elastomeric materials.

%All the systems behave following an exponential behaviour after 5$\%$ of the applied stress,  typical response off elastomeric materials. Stress is defined as $F/A$ where $F$ is the force and $A$ is the area on which the force was applied. Strain is defined as change in length after the application of the stress \textcolor{red}{should I put more math?} Systems with $D_6d$ symmetries, i.e. more oblate structures, have higher toughness compared to the other MPs. That mean that these systems can absorb more energy.Add how the stress was calculated and how the strain was calculated.

To further quantify the elasticity and response of the 
    developed metaparticles to stress, we evaluated their toughness, $\text{T}$,
%Toughness represents the energy of mechanical deformation per unit volume before fracture. 
    as the integration of the stress-strain curve (\Fig{Fig:stress}B). 
%Toughness is mathematically defined as: $\text{T} = \int_0^{\varepsilon_f} \stress \, d\varepsilon$ where $\varepsilon$ is strain, $\varepsilon_f$ is the strain upon failure and $\stress$ is stress. 
The highest toughness is measured by the MPs 
    with D$_\textrm{6d}$ symmetry, with
    values in the 50-60 $\varepsilon/ \sigma^3$ interval. 
Their oblate conformation as well as the direction of 
    applied stress, \ie along the short symmetry axis,
    allows the accommodation of higher mechanical deformation.
This may explain the lack of dependence of toughness on
   metaparticle size.
The small MPs ($\Rg<$ 3~$\nm$) have
    comparable toughness independently of topology.
For the larger metaparticles, particularly among those with 
    icosahedral symmetry, {\MPoneeighty} shows the lowest
    resistance to stress and {\MPtwoforty} the highest,
    suggesting that size plays a secondary role in defining
    the amount of energy MPs can store. 
The response to stress varies also with the location 
    of the applied stress, \ie the area of applied deformation or the arrangement of the pentagons
    relative to the pulling hexagon.
%\textcolor{red}{It's important to notice that the specific zone on which the stress is applied has a role in how the MPs respond to stress. In {\MPoneeighty} case, the stress forces are applied to an hexagon that is adjacent to an isolated pentagon. This characteristic might have an effect on how the entire MP is resisting and handling the applied stress. }

%\textcolor{red}{do you see any differences between pentagons and hexagons? like average energy per pentagon higher than per hexagon?}\textcolor{blue}{Should be done for all systems probably}
\begin{figure*}[ht!]
\includegraphics[width=0.7\textwidth]{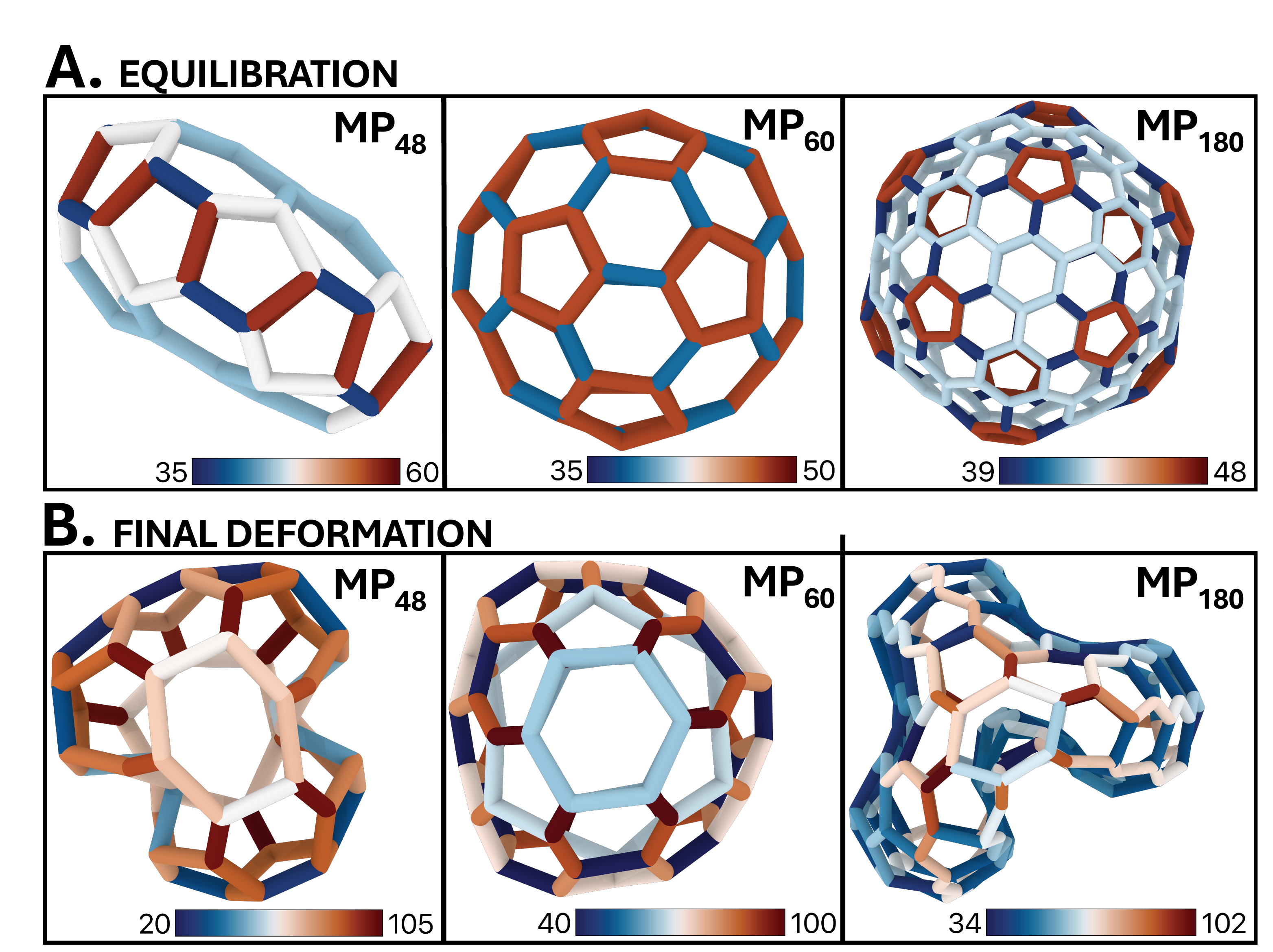}% Here is how to import EPS art
\caption{Average bond energy distribution for {\MPfortyeight}, {\MPsixty}, {\MPoneeighty} in different simulation section. Legend for the energy values represents the average energy for all the bonds for each specific system. (A) Average energy per bond calculated over the whole equilibration run. On average bonds that are part of pentagons have higher energy compared to the ones part of hexagons. (B) Average energy per bond calculated over the whole pulled run.  %\textcolor{red}{On average the energy per bond is higher for smaller MPs since all the bonds will take part to the stress-response.}
\label{Fig:enbond}}
\end{figure*}

To understand this effect, we evaluated the average pulling area 
    ({\Apull}) 
    at the ultimate tensile strength (UTS), \ie at the maximum
    applied stress a system can handle prior to bond breakage (\Fig{Fig:stress}C). 
Two different regions can be identified depending on 
    {\Apull}.
The stress applied to areas of $\approx$ 7.8$\sigma$
    as in the case of {\MPtwenty}, {\MPseventytwo}, 
    {\MPtwoforty}), leads to 
    higher resistance with UTS values over 200 $\varepsilon / \sigma^3$.
Notable, is that the largest and the smallest MPs, 
    \ie {\MPtwoforty} and {\MPtwenty}, respectively, 
    withstand the highest applied stress.
More variability is observed among the metaparticles
    with the stress subjected to areas of almost
    twice the size.
Among them, {\MPoneeighty} has the lowest UTS, 
    which may contribute to the reduced toughness of the
     metaparticle and potentially relate to the propagation
    of stress along its surface.
Hence, our results show a correlation between the average pulling 
    area and the ability of the MPs to respond to the applied stress.
This correlation is independently of size and topoloogy.

%%%%%%%%%%%%%%%%%%%%%%%%%%%%%%%%%%%%%%%%%%%
\subsection{Stress propagation}
To understand how the induced deformation propagates along the 
    surfaces of the  metaparticles, we evaluated the energy distribution
    per bond at different stages during the deformation.
Briefly, we calculated the average energy per bond as the mean over 
    the unconstrained runs and over the pulling simulations, evaluated 
    at the highest strain values (\Fig{Fig:enbond}).
The analysis focused on the systems devoid 
    of external forces 
    reveals that generally 
    the overall energy per pentagon is constant, while
    the energy is distributed differently per bond
    depending on the neighboring polygon
    (\Fig{Fig:enbond}A).
For instance, the energy per pentagon bond in {\MPfortyeight}, 
    is higher when the edge is shared with another pentagon, than
    when it is shared with a hexagon.
By comparison, the energy per bond is homogeneously distributed
    in the systems where a pentagon is surrounded only
    by hexagons 
    (e.g.\Fig{Fig:enbond}A for {\MPsixty} and {\MPoneeighty}).
Similarly, isolated hexagons have a homogenous 
    energy distribution over their bonds.
All in all, isolated pentagons, store, on average, 
    higher energies per bond than the hexagons.
When considering the whole polygons, the energies
    become comparable.
    
During the initial steps of the applied deformation, the 
    strain reflects into an energy increase on the
    bonds neighboring the stress application region (Fig. S2-11). 
In particular, the energy of the bonds aligned nearly in 
    parallel with the direction of the exerted stress (dark red)
    is higher than that of the transversal bonds (dark blue).
This behavior is preserved throughout the pulling simulations and the
    results show that at the maximum extension,
    the bond energies increase symmetrically with 
    increasing distance from the center of the MP, with
    the bonds closest to the pulling extremes having 
    the highest energies (\Fig{Fig:enbond}C).
Hence, the distribution of stress over the bonds is 
    heterogeneous and varies with system size.
Given a  metaparticle with three links ber bead, 
    the distribution of stress is done over a maximum of $3n/2$ bonds, 
    where n is the number of beads.
%This contributes to the maximum axial strain that MPs can
%    handle, which is generally lower for \textcolor{red}{smaller ones.} (\Fig{Fig:stress}) - I don't understand this sentence.

To minimize the average energy per bond under high stress 
    conditions, the metaparticles 
    undergo internal rearrangements.
The rearrangements can be spontaneous, \ie in one step,
    or sequential, \ie in two or multiple steps. 
The small MPs ($\Rg< 3\sigma$) deform in one step
    along the direction of the applied stress
    independently of their symmetry.
In this case, the excluded volume interactions between the
    beads prevent the system from undergoing any
    internal deformations.
For the larger and systems with $D_\textrm{6d}$ symmetry, 
    the topology largely influences
    which deformation mechanism the  metaparticles follow.
For instance, {\MPoneeighty} and {\MPtwoforty} follow a
    multi step deformation. 
In the first steps of the deformation, the MP extends
    along the direction of the pull.
Then, the MP contracts as a response 
    and attains a flower-like conformation 
    determined by the
    arrangement of the pentagons (Fig. S8).
%    where aligned (with respect of the pulling direction) presence of pentagons or alternation of hexagon-pentagon-hexagon drives the creation of unique re-arrangements. 
Next, under continued applied stress, the complex rearranges 
    and bends
    inward upon itself as response to the applied forces.
This arrangement is preserved throughout the simulation
    or until the maximum extension
    limit has been reached.
As eluded, the rearrangement depends on the topology of the 
    MP and the arrangement of the pentagons
    relative to the direction of the pull due to
    the anisotropy in surface tension.
%\textcolor{red}{On average the energy per bond is higher for smaller MPs since all the bonds will take part to the stress-response, while, in larger system, only the bonds closer to the pulling polygon play a major role in stress response. }
%This anisotropy reflects also in the average area per pentagon.

%%%%%%%%%%%%%%%%%%%%%%%%%%%%%%%%%%%%%%%%%%%%%%%%%%%%%%%%%
\subsection{Deformation and recovery}
\begin{figure}
\includegraphics[width=\columnwidth]{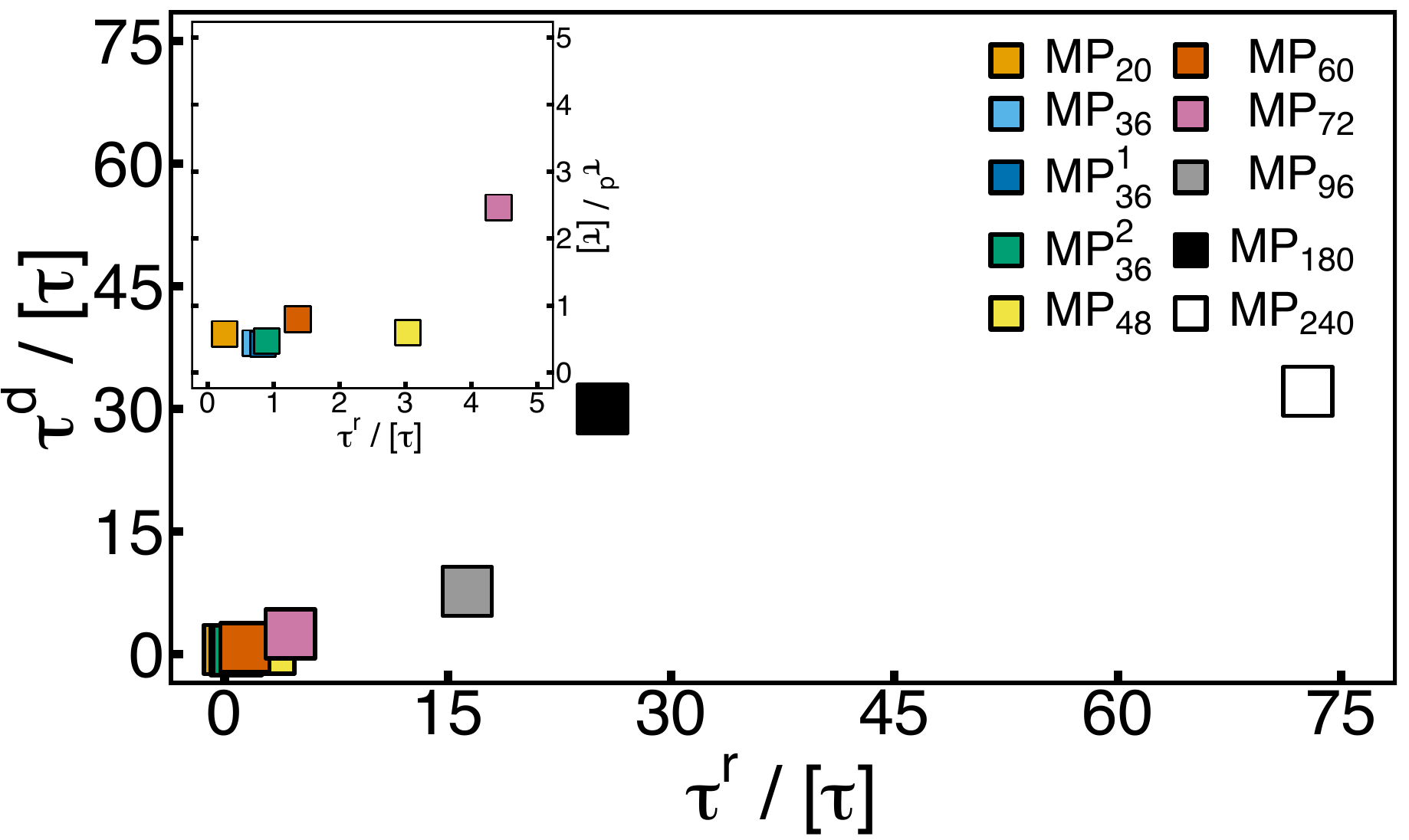}% Here is how to import EPS art
\caption{\label{Fig:times} Deformation ($\tpull$) 
    and recovery times ($\trelax$).
    The insert shows the behavior of the small MPs.}
\end{figure}
Next, we investigated the ability of the metaparticles to recover to their
    initial conformation upon stress release.
For this, we compared the time required for the MPs 
    to reach their maximum extensibility (evaluated at the UTS),
    $\tpull$, with the recovery time, $\trelax$, 
    \ie the time required for the MPs to
    return to their initial conformation after stress release
    (\Fig{Fig:times}).
A system is considered as fully recovered, when the transversal and 
    axial diameters are within the standard deviation of their
    respective initial equilibrated values. 
%    \textcolor{red}{The relaxation phase is calculated over a variable simulation time dependent on the time required to each specific MP to get back within it's original diameter. These time are: 5 $\tau$ for {\MPtwenty}, {\MPthirtysix},{\MPthirtysixone},{\MPthirtysixtwo} and {\MPsixty}, 10 $\tau$ for {\MPfortyeight}, 20 $\tau$ for {\MPseventytwo} and 100 $\tau$ for {\MPnintysix}, {\MPoneeighty} and {\MPtwoforty}. All simulation where performed with a $10^{-5}$ timestep in order to better catch the relaxation phase.}
The analysis reveals that the deformation time increases with 
     metaparticle size, except for {\MPoneeighty} and {\MPtwoforty}, 
    which have comparable $\tpull$. 
The longer $\tpull$ may be associated with the 
    deformation mechanism, as MPs undergoing multistep deformations
    require longer times to reach their maximum extension.
    
The comparison between the deformation and recovery times shows 
    a linear correlation for
    small MPs ($\Rg < 4\sigma$), which
    require comparable times to deform 
    and to recover
    independently of their topology (\Fig{Fig:times}). 
The relaxation time of {\MPfortyeight} is about three times longer
    than its deformation time due to the direction of applied
    stress, \ie the forces are applied along the shortest
    axis due to symmetry reasons.
With increasing system size, 
    topology and  metaparticle deformation affect differently 
    the two timescales.
For instance, the deformation and internal rearrangement of {\MPtwoforty}
    reflects in the longer $\trelax$ as compared to $\tpull$ 
    (72.8 $\tau$ and 32.1 $\tau$, respectively). 
The longer $\trelax$ for {\MPtwoforty} can be linked to the 
    larger number of bonds that are involved in the relaxation 
    process as compared to the other systems.
Furthermore, visual inspection revealed that the complete
    recovery to the initial state is limited by the 
    conformation of a pentagon, which remains in the
    "inward" folded configuration.
This unique conformation reflects minimally on the overall
    conformation of the MP.
Elongated runs revealed that the complete recovery of this
    pentagon extends on 100-fold longer timescales.
% We performed a longer simulation of $10^4$ $\tau$ and we noticed that the full release of the pentagon occur after $\approx$ 7480 $\tau$. The entire {\MPtwoforty} structure is relaxed following our ruling, but for larger systems that {\MPtwoforty}, some unique deformations events might be visible for longer times.
    
Similar to the multi-step deformation, the recovery of 
    the  metaparticles can also occur following different
    pathways, \ie single or multi-step relaxation.
On the one hand, the relaxation of MPs that 
    do not undergo internal rearrangements
    occurs in a single step, \ie the particle contracts upon stress 
    release. 
The thermal fluctuations and the FENE potential gradually 
    drive the
    MP to its initial conformation.
On the other hand, when MPs undergo internal rearrangements, 
    the system requires a longer recovery times 
    due to the multi-step recovery process. 
For instance, upon stress release 
    {\MPfortyeight} first axially contracts, while maintaining its
    "folded" conformation.
Next, the deformed surfaces recover one by one to their 
    equilibrium state.
The complete recovery to the initial state as 
    response to stress is a property of elastomeric materials, 
    particularly bio-inspired ones \cite{Yoshida2017}.

%In general, as expected, the MPs are stretched and relaxed according to their size with smaller MPs which are faster and larger one which are slower with the only exception of {\MPfortyeight} which relax slower compared to the larger {\MPsixty} due to how the pulling forces were applied along the shorter size of the MP and to its contraction at the UTS.
    
%%%%%%%%%%%%%%%%%%%%%%%%%%%%%%%%%%%%%%%%%%%%%
\section{Discussion and conclusion}
We developed a coarse-grained model for \textit{MetaParticles},
    \ie soft, resilient nanoparticles with tunable 
    and responsive properties.
This was achieved by representing an MP as 
    a collection of beads interconnected by springs. 
The topological arrangement of the beads and the bonds
    is symmetrical and was inspired 
    by the topologies of closed fullerenes.
Hence, each bead is linked to three surrounding beads
    forming a pattern of hexagons and 12 pentagons, endowing the
    the molecules with quasi-spherical shapes. 
This topological arrangement brings unique symmetries 
    to each metaparticle.
Our results show that the MPs are responsive to the applied stress
    following an elastomeric behaviour driven by a combination of 
    size, topology (symmetry) and area of applied stress.
The deformation is reversible as the  metaparticles return
    to their original shapes upon stress release.
%    , characteristic for elastomeric materials.
Furthermore, the induced deformation and the relaxation occur on 
    different timescales depending
    on the deformation and relaxation mechanisms.
%The deformation times of the studied MPs generally increase with
%    system size, apart from the unique case of {\MPoneeighty}.
Single step deformation leads to a linear correlation 
    between the two timescales and is typical for 
    MPs with $\Rg<$ 3~$\nm$.
Multistep deformations and relaxations require 
    longer times due
    to the internal rearrangements of the beads, a feature shared
    by the larger metaparticles.
    
As discussed in the Introduction, 
    the development of a 
    particle with tunable and responsive
    properties constitutes
    a major step towards modeling and understanding 
    the behavior of deformable 
    nanoparticles in the biological environment.
Particularly, our aim is to introduce a generic model,
    easy to fine tune to include specificity, to be used
    to understand the responsiveness and adaptability of semi-flexible nanocarriers 
    to environmental, chemical and physical 
    changes present in biological systems.
We envision a model, in which the beads of the  \textit{MetaParticles}
    represent a collection of ligands on the surface of
    a nanocarrier.
These beads will be able to adjust their 
    interactions in response to the environment,
    e.g. they can become attractive or repulsive depending
    on the proximity to or embedding in the membrane, hence capturing the
    specific properties of compounds at different pH.
These interactions are to be extracted from atomistic simulations
    between specific ligands and the cellular membrane.
As highlighted throughout this manuscript, the generic behavior
    of this model extends beyond pure drug delivery understanding
    and aids in the design of metamaterials with bioinspired properties \cite{Ma2020}.
Future questions will address the 
    effects of topology, defects and activity 
    on the mechanical properties
    of the  metaparticles.

\begin{acknowledgments}
 I.M.I. acknowledges support from the Sectorplan B\`{e}ta \& Techniek of the Dutch Government and the Dementia Research - Synapsis Foundation Switzerland. 
\end{acknowledgments}
%\nocite{*} 
%\bibliography{aipsamp.bib}% Produces the bibliography via BibTeX.


\end{document}

%\preprint{Paesani}

\title[Computational engineering of responsive metaparticles]{
Supplementary information \\ 
MetaParticles:
Computationally engineered nanomaterials with tunable and responsive properties}
% Force line breaks with \\
\author{Massimiliano Paesani$^{1,2,3}$}
\author{Ioana M. Ilie$^{1,2,3}$}%
 \email[Corresponding author: ]{i.m.ilie@uva.nl}
\affiliation{ 
$^1$ Van 't Hoff Institute for Molecular Sciences, University of Amsterdam, Amsterdam,
 The Netherlands \\ 
$^2$ Amsterdam Center for Multiscale Modeling (ACMM), University of Amsterdam, the Netherlands \\
$^3$ Computational Soft Matter (CSM), University of Amsterdam, the Netherlands
}%Lines break automatically or can be forced with \\
\maketitle

%%%%%%%%%%%%%%%%%%%%%%%%%%%%%%%%%%%%%%%%%%%%%%%%%%%%%%%

\begin{figure}
\includegraphics[width=\columnwidth]{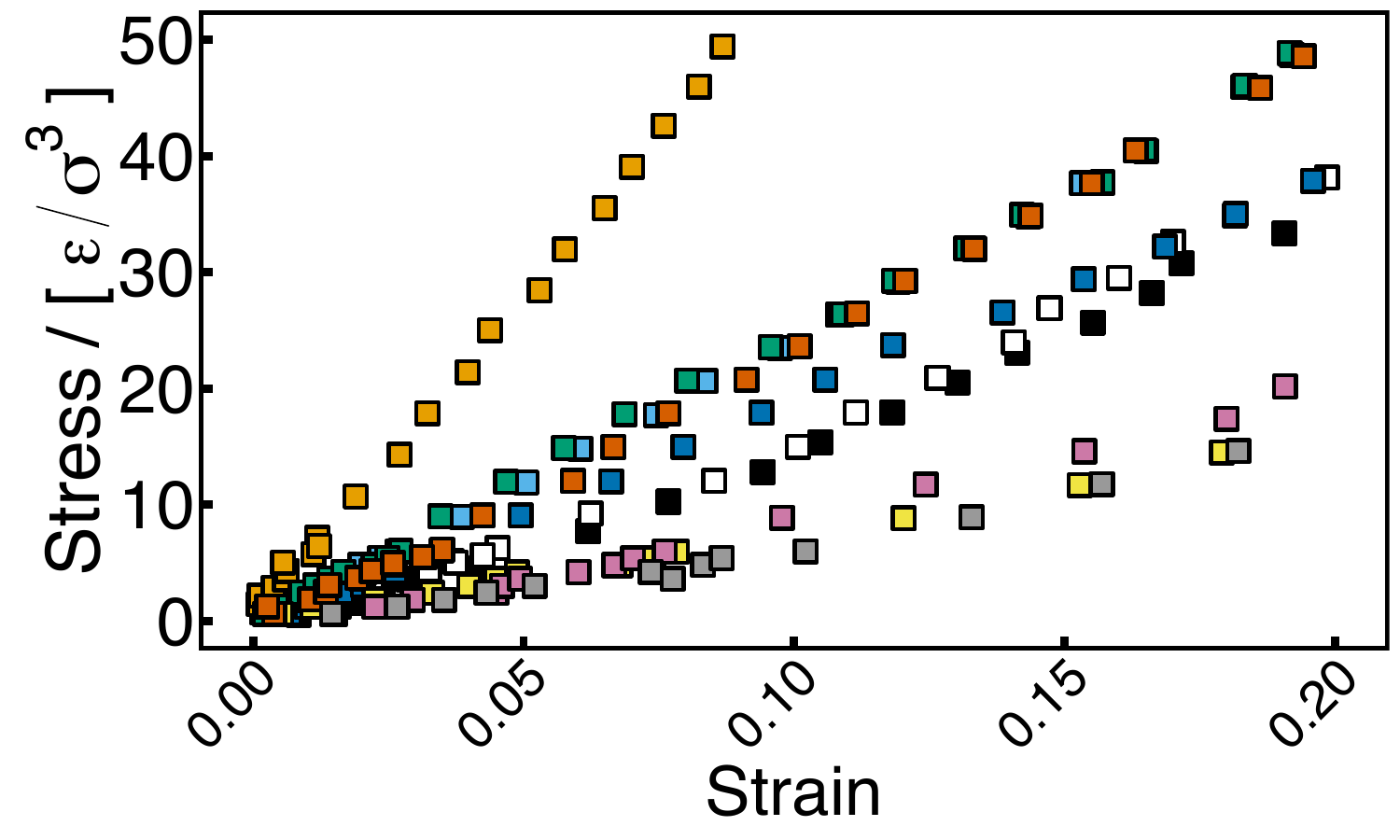}% Here is how to import EPS art
\caption{ \textbf{Mechanical properties of the metaparticles.} Stress-strain curves zoom-in extracted from Fig. 3A.}
\end{figure}

\begin{figure}
\includegraphics[width=\columnwidth]{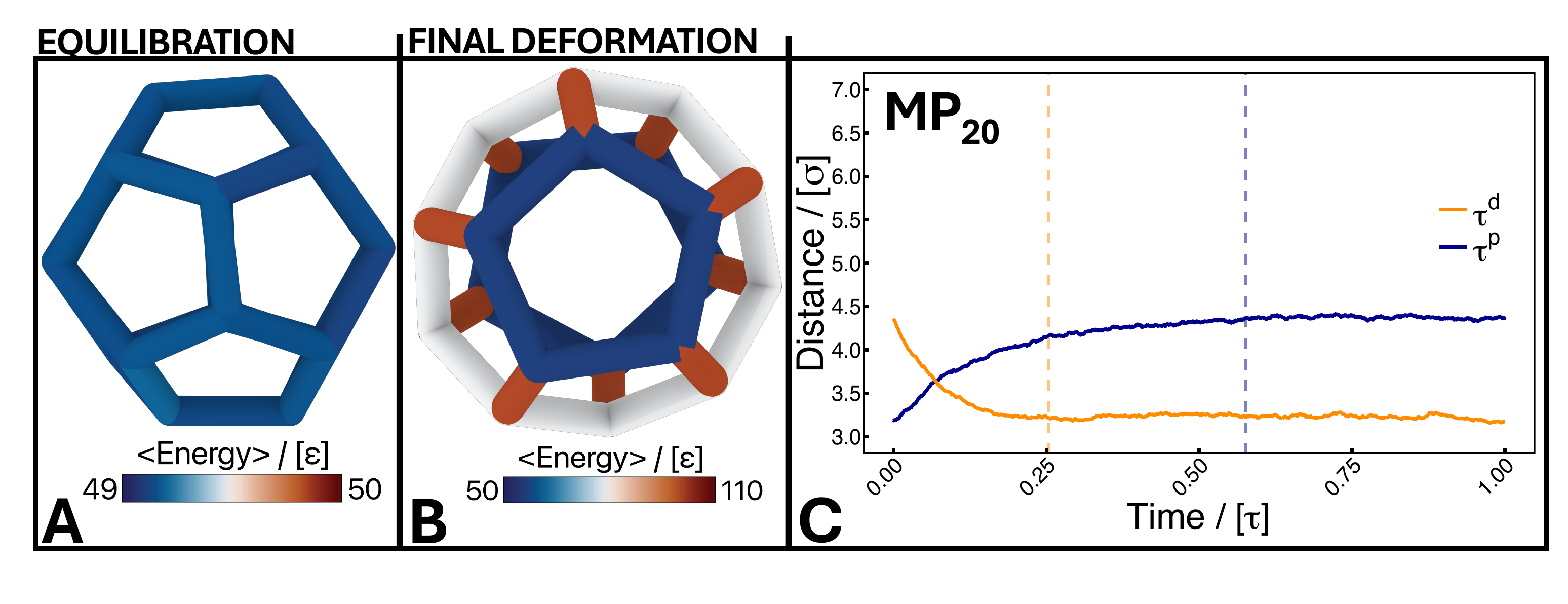}% Here is how to import EPS art
\caption{\label{fig:epsart}  Average bond energy during the equilibration simulation  (A). Average bond energy during the extension simulation. The snapshot is a top view of the MP  aligned along its deformation axis (B). Pulling and relaxation timelines ($\tpull$ and $\trelax$) in blue and orange, respectively (C). The vertical dashed lines represent the times beyond which no additional deformation of the particle is considered. Each individual time has been calculated as the first value that is within the standard deviation of the last 30,000 steps of the simulation.
They correspond to the values shown in Fig. 5.}
\end{figure}

\begin{figure}
\includegraphics[width=\columnwidth]{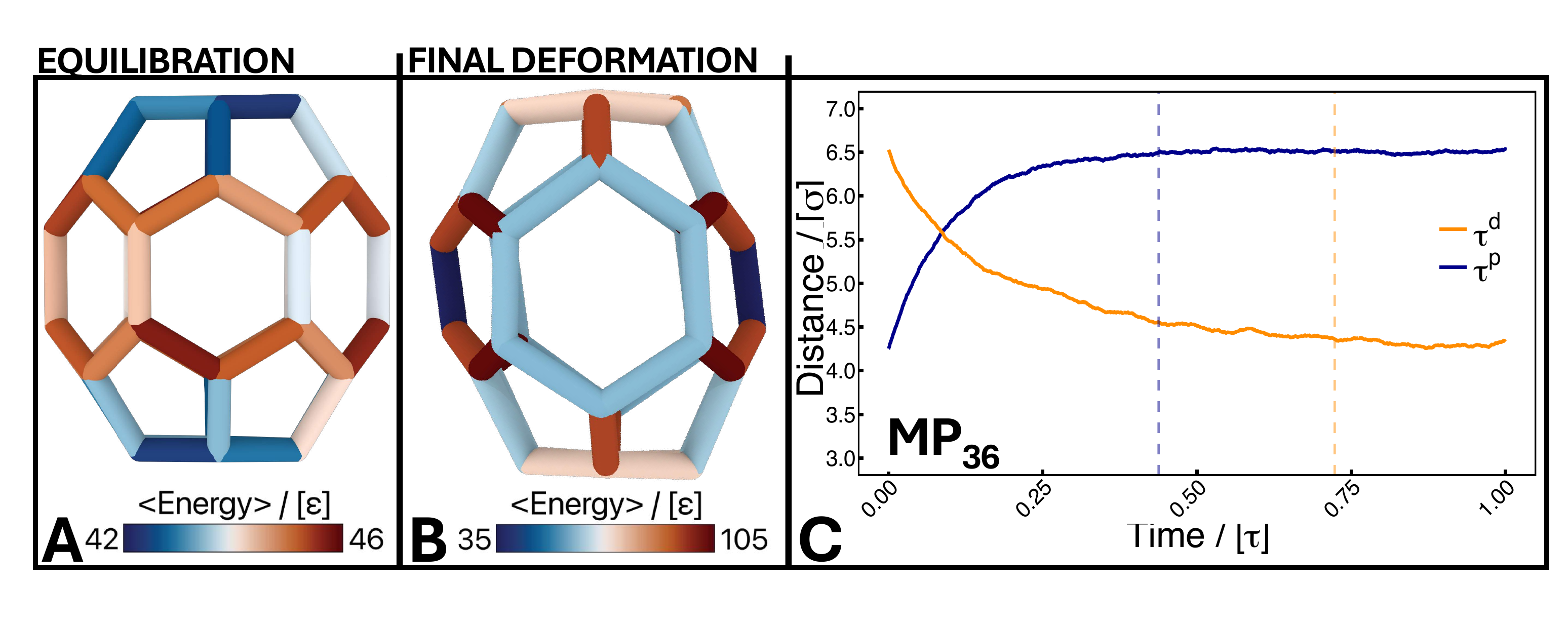}% Here is how to import EPS art
\caption{See \Fig{fig:epsart}.}
\end{figure}

\begin{figure}
\includegraphics[width=\columnwidth]{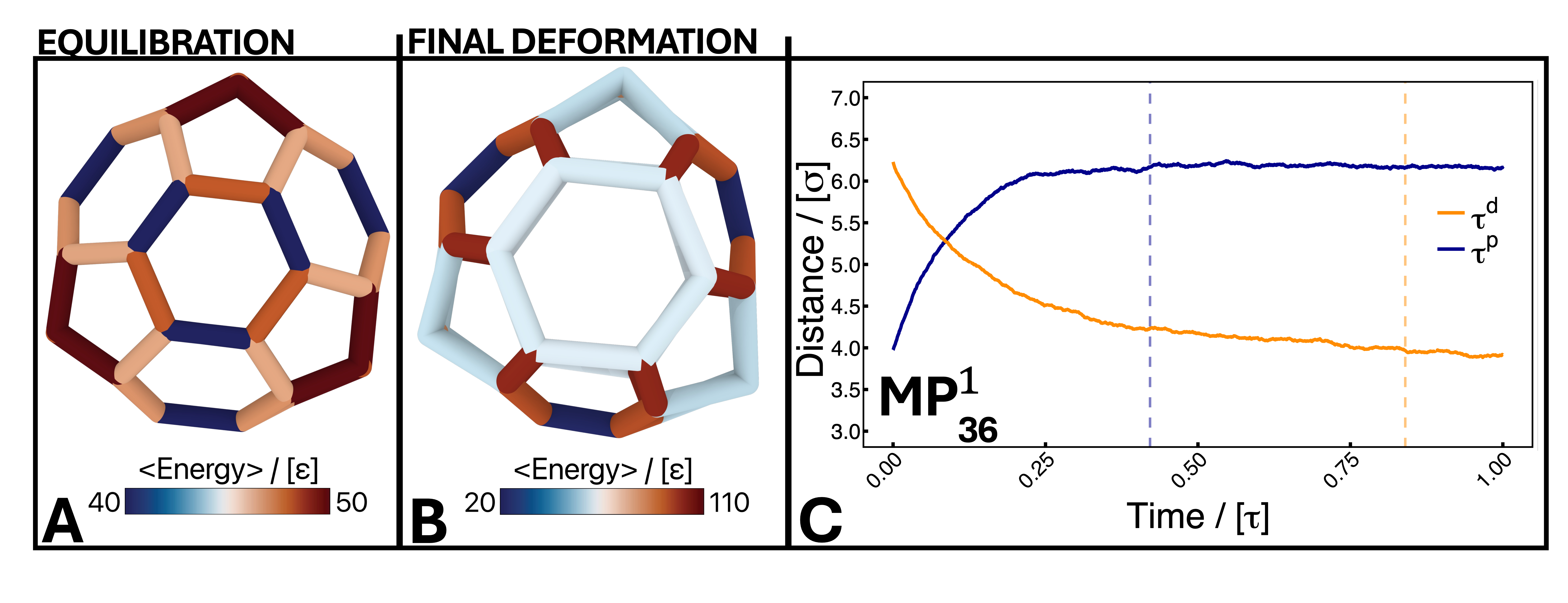}% Here is how to import EPS art
\caption{See \Fig{fig:epsart}..}
\end{figure}

\begin{figure}
\includegraphics[width=\columnwidth]{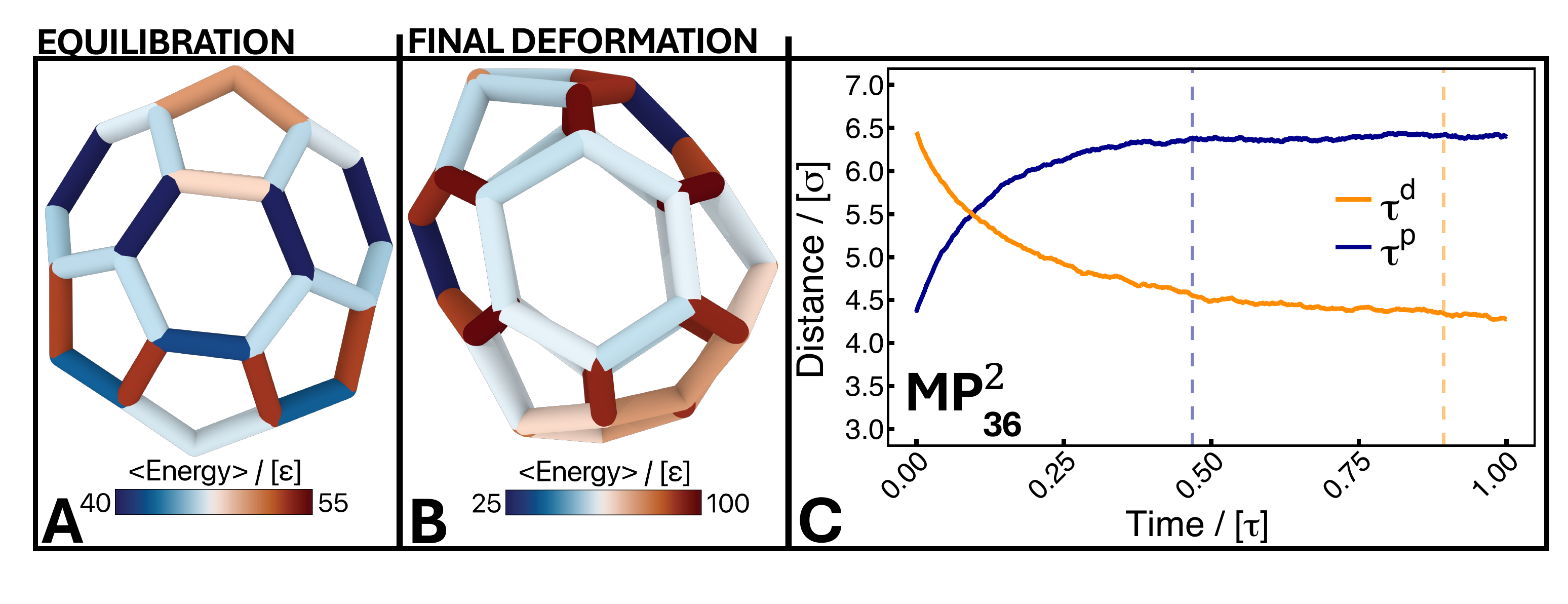}% Here is how to import EPS art
\caption{See \Fig{fig:epsart}.}
\end{figure}

\begin{figure}
\includegraphics[width=\columnwidth]{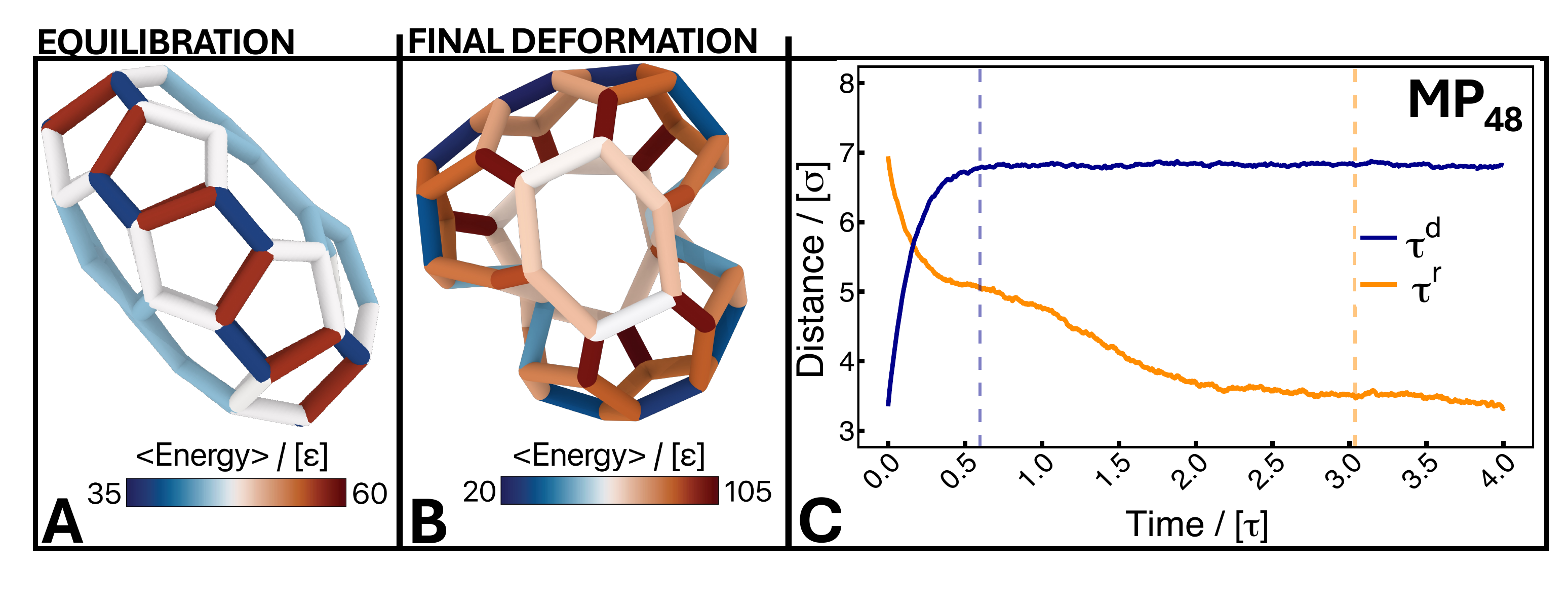}% Here is how to import EPS art
\caption{See \Fig{fig:epsart}.}
\end{figure}

\begin{figure}
\includegraphics[width=\columnwidth]{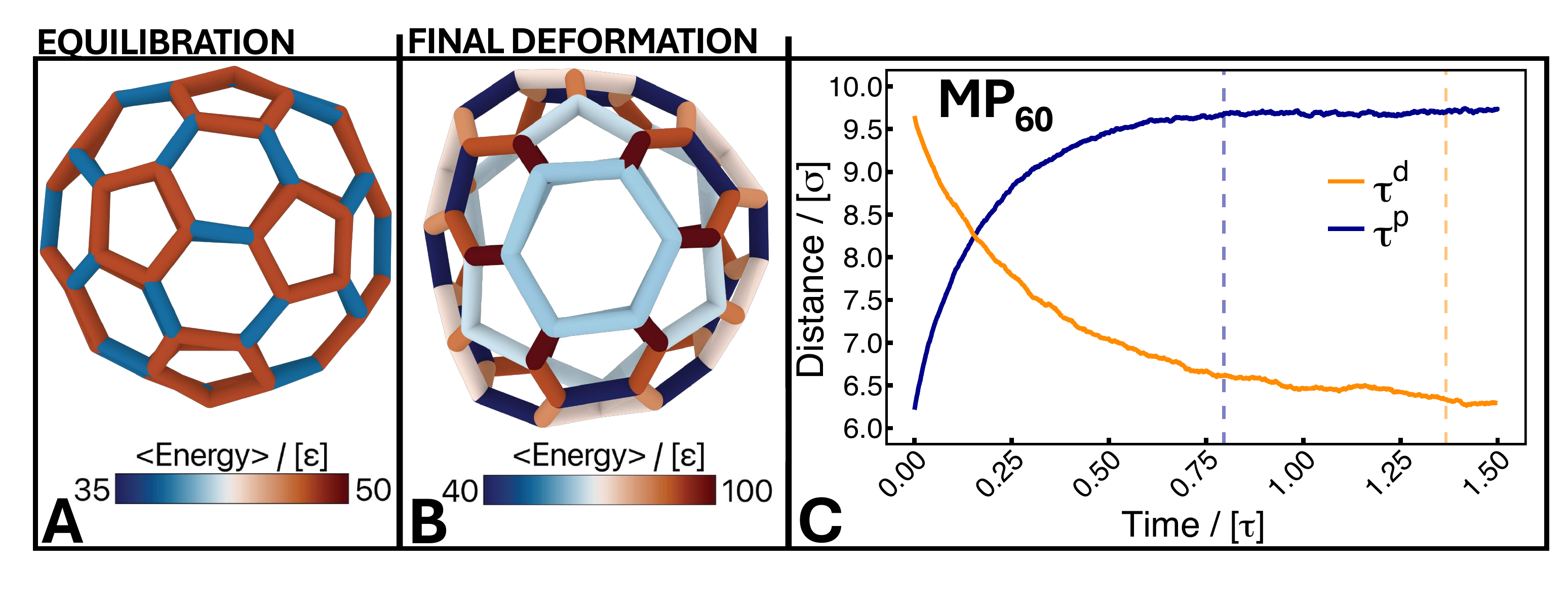}% Here is how to import EPS art
\caption{See \Fig{fig:epsart}.}
\end{figure}

\begin{figure}
\includegraphics[width=\columnwidth]{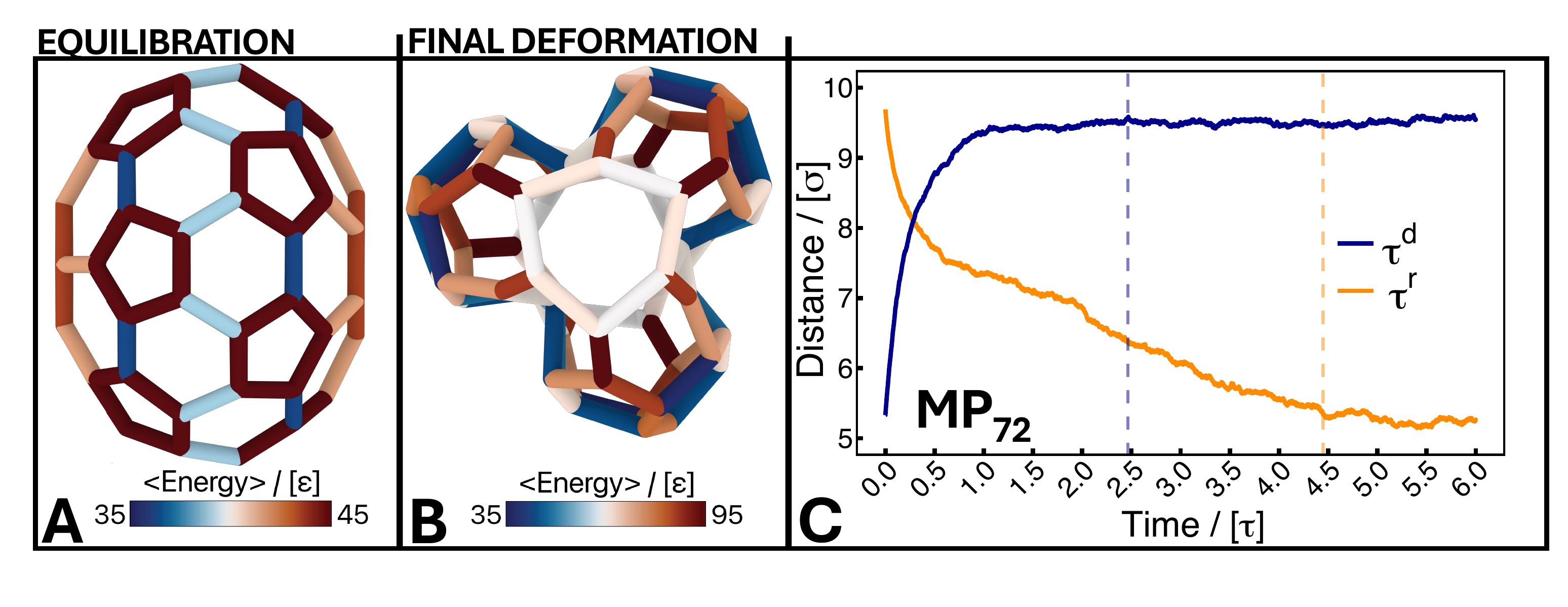}% Here is how to import EPS art
\caption{See \Fig{fig:epsart}.}
\end{figure}

\begin{figure}
\includegraphics[width=\columnwidth]{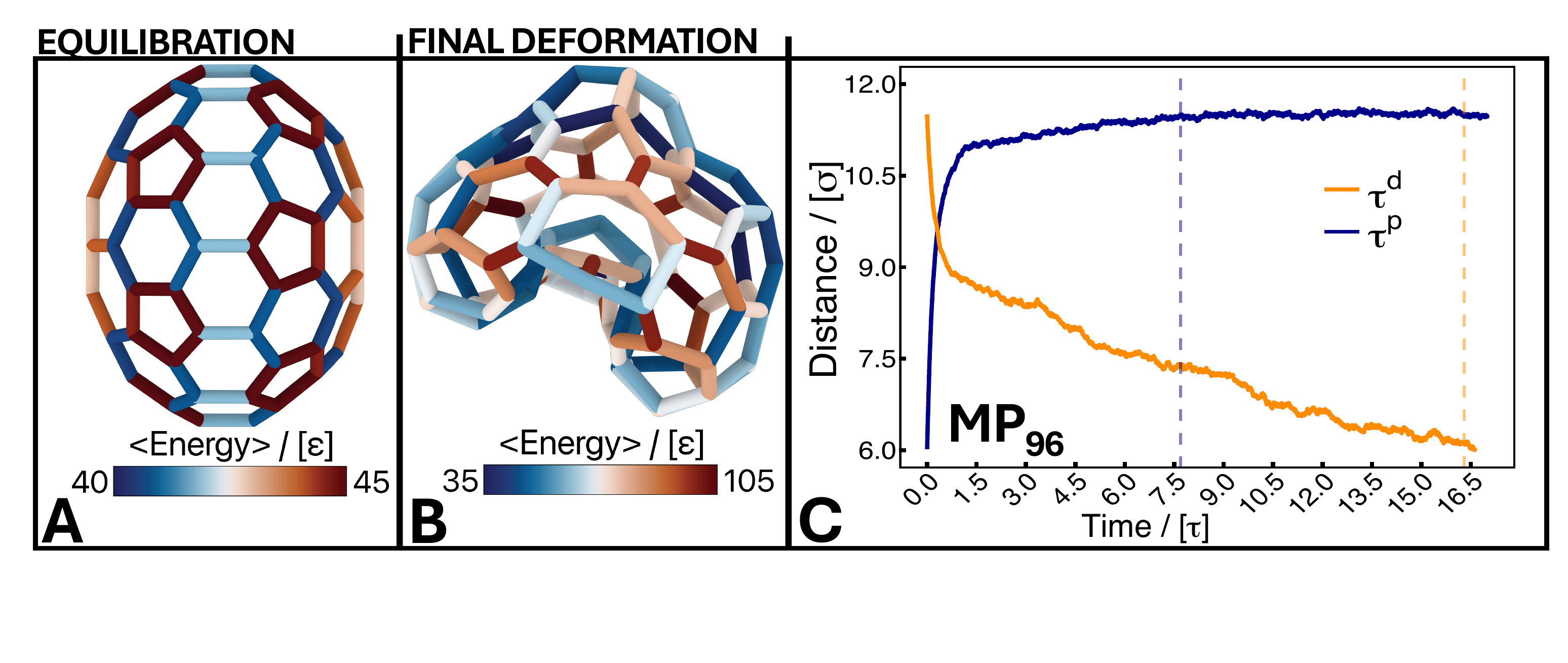}% Here is how to import EPS art
\caption{See \Fig{fig:epsart}.}
\end{figure}

\begin{figure}
\includegraphics[width=\columnwidth]{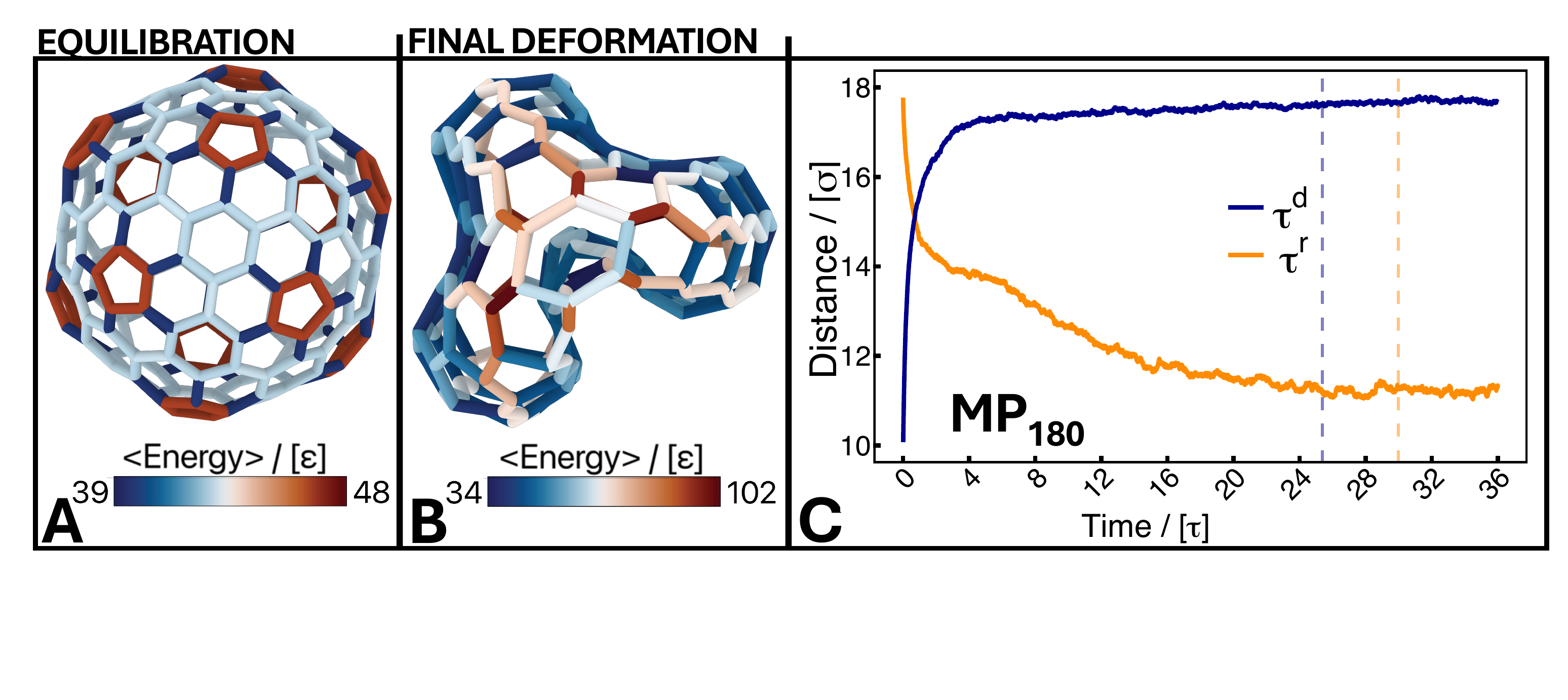}% Here is how to import EPS art
\caption{See \Fig{fig:epsart}.}
\end{figure}

\begin{figure}
\includegraphics[width=\columnwidth]{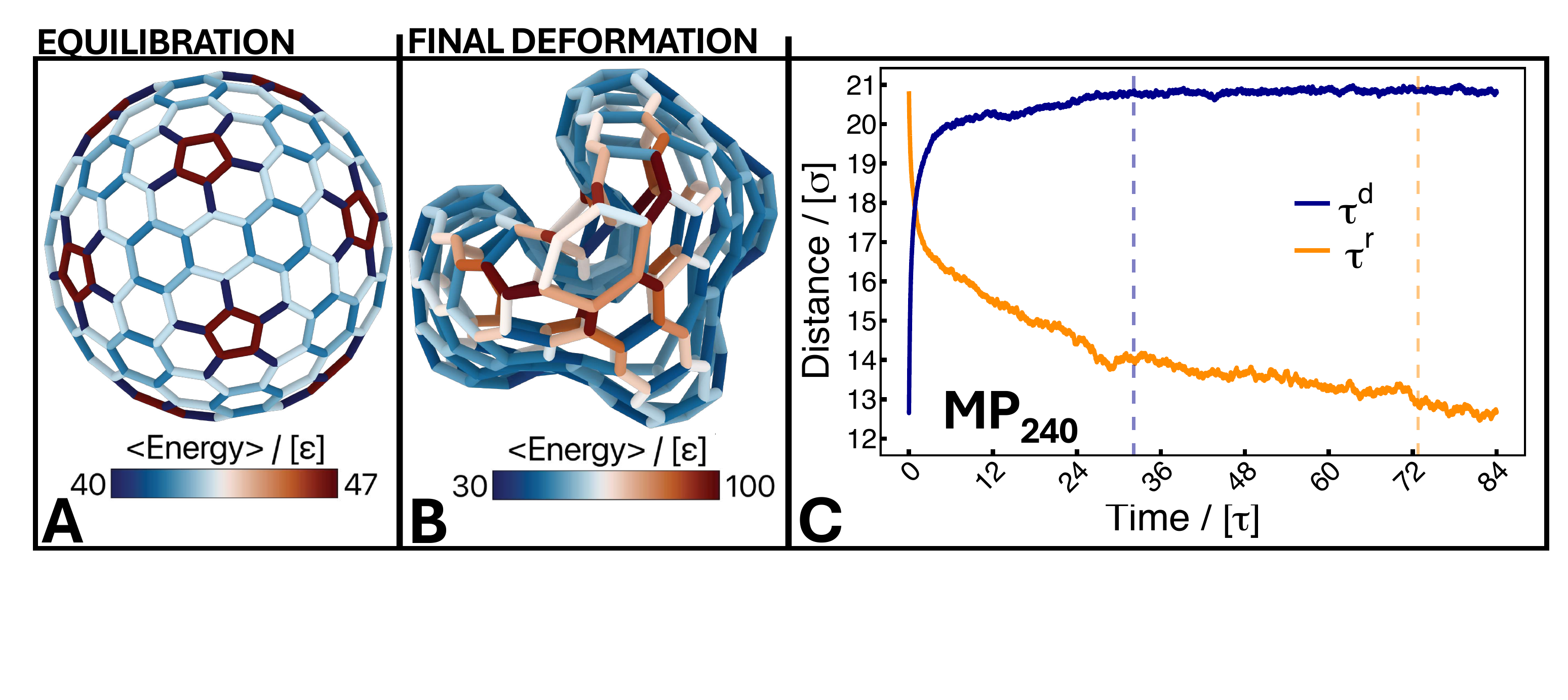}% Here is how to import EPS art
\caption{See \Fig{fig:epsart}.}
\end{figure}

\nocite{*}